%% file: TUGL_v0.tex
\documentclass{article}
\usepackage{spconf}
\usepackage{graphicx}
\usepackage{amsmath}
\usepackage{amssymb}

\usepackage[colorlinks]{hyperref}
\usepackage{verbatim}
\usepackage{color}

\input{defs}

\graphicspath{{figs/}}

\usepackage{tabularx} 
\usepackage{booktabs}   
\usepackage{mathrsfs} 
\usepackage{multirow} 
\usepackage{amsfonts} 
\usepackage{wrapfig}  
\usepackage[misc]{ifsym}
\usepackage{cite}
\usepackage{subcaption}

\title{Topology-Aware Graph Augmentation for
Predicting Clinical Trajectories in
Neurocognitive Disorders}
\name{
Qianqian {Wang}$^1$, 
Wei Wang$^2$, 
Yuqi {Fang}$^1$, 
Hong-Jun Li$^2$,  
Andrea {Bozoki}$^3$, 
Mingxia {Liu}$^{1,*}$
}
\address{$^1$Department of Radiology and BRIC, University of North Carolina at Chapel Hill, NC 27599, USA \\
$^2$Department of Radiology, Beijing Youan Hospital, Capital Medical University, Beijing, China\\
$^3$Department of Neurology, University of North Carolina at Chapel Hill, NC 27599, USA
}
\begin{document}
%
\maketitle
\begin{abstract}
Brain networks/graphs derived from resting-state functional MRI (fMRI) help  study underlying pathophysiology of neurocognitive disorders by measuring neuronal activities in the brain.  
Some studies utilize learning-based methods for brain network analysis, but typically suffer from low model generalizability caused by scarce labeled fMRI data. 
As a notable self-supervised strategy, 
graph contrastive learning helps leverage auxiliary unlabeled data. 
But existing methods generally arbitrarily perturb graph nodes/edges to generate augmented graphs, 
without considering essential topology information of brain networks. 
To this end, we propose a topology-aware graph augmentation (TGA) framework, comprising  
a \emph{pretext model} to train a generalizable encoder on large-scale unlabeled fMRI cohorts and a \emph{task-specific model} to perform downstream tasks 
on a small target dataset. 
In the pretext model, we design two novel topology-aware graph augmentation strategies: 
(1) \emph{hub-preserving node dropping} that 
prioritizes preserving brain hub regions according to node importance, 
and 
(2) \emph{weight-dependent edge removing} that   
focuses on keeping important functional connectivities based on edge weights. 
Experiments on $1,688$ fMRI scans
suggest that TGA outperforms several state-of-the-art methods. 
\end{abstract}
\begin{keywords}
Graph augmentation, Graph topology,  Functional MRI 
\end{keywords}

\section{INTRODUCTION}
\label{sec:intro}

Brain networks/graphs derived from resting-state functional MRI (fMRI) help objectively study underlying pathophysiology of neurocognitive disorders by measuring brain neuronal activities.  
Some fMRI-based studies utilize learning-based approaches for brain network analysis and imaging biomarker identification. 
But existing  methods often suffer from poor model generalizability due to limited labeled data~\cite{cui2022interpretable}.

As a promising self-supervised learning strategy, 
graph contrastive learning provides an effective solution to pretrain a generalizable encoder on auxiliary data without requiring task-specific labels~\cite{chen2021exploring}.
In general, 
it first performs graph data augmentation for each input graph to generate two views, and then optimizes the agreement of graph representations from these two views to achieve self-supervised model training~\cite{ hassani2020contrastive}. 
However, existing studies typically 
randomly perturb graph nodes or edges to generate augmented graphs for contrastive learning~\cite{you2020graph},
ignoring the crucial topological information conveyed in functional brain networks.

\begin{figure*}[!t]
\setlength{\abovecaptionskip}{-1pt} 
\setlength{\belowcaptionskip}{-4pt}  
\setlength\abovedisplayskip{-1pt}
\setlength\belowdisplayskip{-1pt}
\begin{center}
\includegraphics[width=0.96\linewidth]{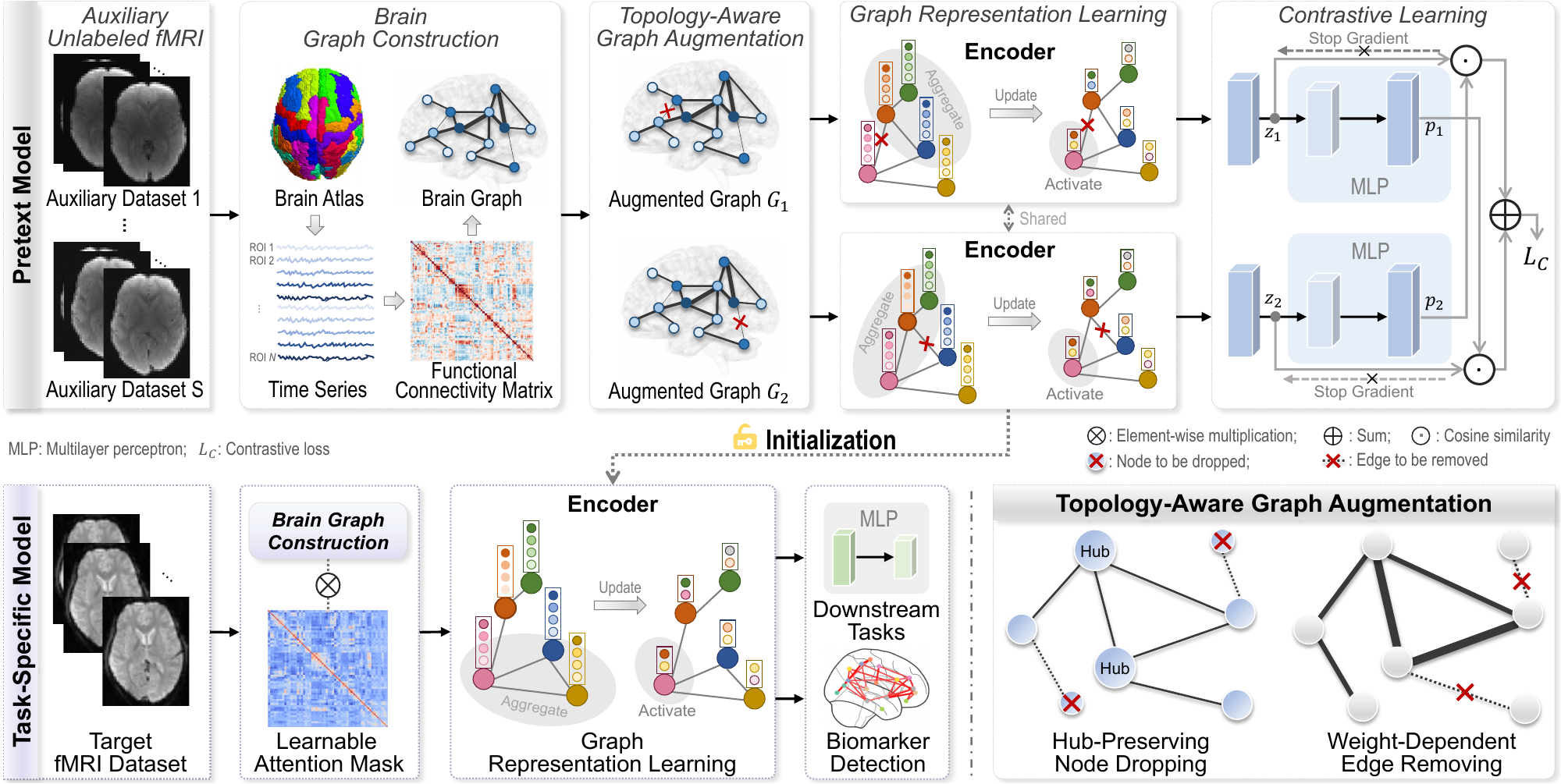}
\end{center}
   \caption{Illustration of the proposed topology-aware graph augmentation (TGA) framework for functional MRI analysis.
}
\label{pipeline}
\end{figure*}

To this end, we propose a topology-aware graph augmentation (TGA) framework for clinical trajectory prediction of HIV-associated neurocognitive disorders (HAND) with resting-state fMRI. 
As shown in Fig.~\ref{pipeline}, the TGA consists of  a \emph{pretext model} to train a generalizable graph convolutional network (GCN) encoder on large-scale unlabeled fMRI cohorts
and a \emph{task-specific model} to fine-tune the encoder for HIV-associated disorder identification and clinical score prediction on a target HAND dataset. 
We design two novel topology-aware graph augmentation strategies
including
(1) \emph{hub-preserving node dropping} that 
prioritizes preserving brain hub regions according to node importance, 
and 
(2) \emph{weight-dependent edge removing} that focuses on keeping important functional connectivities based on edge weights. 
Experiments on $1,688$ fMRI scans 
validate the superiority of our TGA  
in classification and regression tasks.
The TGA incorporates a learnable attention mask  to automatically 
detect HIV-related brain regions and functional
connectivities, providing potential imaging biomarkers for early intervention. 

%

\section{Materials and METHODOLOGY}
\subsection{Subjects and fMRI Preprocessing}

Two publicly available cohorts (ABIDE~\cite{di2014autism} and MDD~\cite{yan2019reduced}) and a private HAND dataset with resting-state fMRI are used. 
From ABIDE and MDD, we select subjects
whose time series length is at least 230, yielding $1, 591$
auxiliary fMRI scans for self-supervised model training. 
The HAND dataset is acquired from a local hospital, including 68 patients with HIV who exhibit asymptomatic neurocognitive impairment (ANI) 
and 69 age-matched healthy controls (HC). 
All fMRI scans are preprocessed using a  standard pipeline~\cite{yan2010dparsf}. 

\subsection{Proposed Method}

\vspace{-4pt}
\subsubsection{Pretext Model}
\vspace{-4pt}
Learning-based models generally suffer from poor generalization since labeled fMRI data is limited~\cite{cui2022interpretable}. 
Inspired by~\cite{you2020graph}, 
we design a \emph{self-supervised pretext model} by leveraging large-scale unlabeled fMRI data to train a generalizable encoder that 
can well adapt to downstream tasks. 
As shown in the top panel of Fig.~\ref{pipeline}, 
with auxiliary unlabeled fMRI as input,
the pretext model contains the following 4 components.

\vspace{2pt}
\noindent \textbf{(1) Brain Graph Construction}. 
A brain network derived from resting-state fMRI of a subject can be treated as 
a graph {\small{$G$$=$$\{V,E\}$}}, where {\small{$V$$=$$\{v_{i}\}_{i=1}^{N}$}} is a set of ROIs/nodes, {\small{$E$$=$$\{e_{i,j}\}_{i,j=1}^{N}$}} is a set of functional connectivities/edges, and $N$ is the number of ROIs. 
We define the preprocessed fMRI time series as {\small{$S \in \mathbb{R}^{T\times N}$}}, where {\small{$T$}} is the length of time series. 
For each graph, we generate an adjacency matrix {\small{$A$ $\in$ $\mathbb{R}^{N\times N}$}} by computing Pearson's correlation (PC) between time series of paired ROIs. 
In this work, we denote node feature matrix as {\small{$X=A$}}, where the initial feature is connectivity strength between a node and all other nodes.

\vspace{2pt}
\noindent \textbf{(2) Topology-Aware Graph Augmentation}. 
To facilitate self-supervised training of the pretext model, we design two {topology-aware graph augmentation} methods that preserve graph topology information,
driven by the prior that topological structure is crucial for analyzing functional pathology.

{\emph{Hub-preserving Node Dropping}} 
(HND), which prioritizes preserving brain hub structure according to node importance. 
Here, we use degree centrality (DC) $d_{i}$ to measure the importance of node $v_{i}$, and  
assign dropping probability to each node based on its DC. 
To prioritize preserving hub regions, nodes with higher DC would have lower dropping probability.   
That is, the node dropping probability is inversely proportional to its centrality. 
The probability of a node $v_{i}$ being dropped can be 
defined as {\small{$q_{i}$$=$$1/d_i$}}. 
To obtain the probability distribution for HND, 
we perform probability normalization, represented as {\small{$p_{i}$$=$$q_{i}/{\sum_{i=1}^{N}q_{i}}$}}. 
Finally,
we drop a certain proportion/ratio (\ie, $\alpha$) of nodes following this probability distribution to generate two augmented graphs ${\small{G_{i}}}$ ($i=1,2$). 


{\emph{Weight-dependent Edge Removing}} 
(WER), which focuses on keeping important functional connectivities based on edge weights. 
Similar to HND, 
we define the probability for an edge $e_{i,j}$ being removed as 
{\small{$p_{i,j}$$=$${q_{i,j}}/{\sum_{i=1}^{N}\sum_{i=j}^{N}q_{i,j}}$}}, where {\small{$q_{i,j}$$=$${1}/{a_{i,j}}$ and $a_{i,j}$}}  is the edge weight between node $v_{i}$ and $v_{j}$. 
Here, we employ PC to describe functional connectivity strength (\ie, edge weight). 
Likewise, we remove a certain proportion (\ie, $\beta$) of edges to generate augmented graphs ${\small{G_{i}}}$ ($i=1,2$) following this probability distribution for  edge set. 
In this work, the node dropping ratio $\alpha$ and edge removing ratio $\beta$
are empirically set as $10\%$ and $50\%$, respectively. %

\vspace{2pt}
\noindent \textbf{(3) Graph Representation Learning}.
Considering the superiority of GCN in fMRI feature learning,
the augmented graphs $G_{1}$ and $G_{2}$ are fed into two shared GCN encoders (two GCN layers with hidden dimension of $64$ for each encoder) that can aggregate and update neighboring node features to generate new representations, described as: 
\vspace{-2pt}
\begin{equation}
\small
\begin{aligned}
Z = f(A,X) = \sigma(\hat{A}\sigma(\hat{A}XW^{0})W^{1}),
\end{aligned}
\label{Eq1}
\end{equation}
\vspace{-2pt}where
{\small{$\hat{A} = D^{-\frac{1}{2}}AD^{-\frac{1}{2}}$}} is a normalized adjacency matrix, $\sigma$ is {\small{$ReLU$}}, {\small{$W^{0}$}} and {\small{$W^{1}$}} are weight matrices, and  {\small{$D$}} is a degree matrix.
We then
average over all rows of node feature matrix {\small{$Z$}}
to generate a graph-level representation $z_i$ ($i=1,2$).

\vspace{2pt}
\noindent \textbf{(4) Contrastive Learning}. 
With $z_{1}$ and $z_{2}$ 
 as input, 
we further apply multilayer perceptron (MLP) for feature abstraction, yielding two new representation vectors $p_{1}$ and $p_{2}$.
Following contrastive learning paradigms~\cite{chen2021exploring}, 
our \emph{pretext model} is updated by maximizing the similarity between two augmentation views from the same subject. 
Mathematically, the graph contrastive loss in our TGA can be formulated as:
\vspace{-2pt}
\begin{equation}
\small
\begin{aligned}
\mathcal{L}_{C}=\Psi(\psi(z_{1}),p_2)
+\Psi(\psi(z_{2}),p_1),
\end{aligned}
\label{Eq2}
\end{equation}
\vspace{-2pt}
where $\Psi$ is the negative cosine similarity.
And $\psi$ denotes the stop-gradient operation to ensure that 
the GCN encoder on {\small{$G_{1}$}} receives no gradient from $z_{1}$ in the first term, but it receives gradients from $p_{1}$ in the second term (and vice versa for {\small{$G_{2}$}}),  helping prevent model collapse~\cite{chen2021exploring}. 

\vspace{-4pt}
\subsubsection{Task-Specific Model} 
\vspace{-4pt}
Our goal is to train a generalizable GCN encoder that can well adapt to the task-specific model. 
In the task-specific model, we first construct brain graphs based on target fMRI, and then feed these graphs into a GCN encoder for graph representation learning, followed by an MLP for prediction. 
Note that this encoder is initialized using parameters learned in the \emph{pretext model}. 
In this work, we perform HIV-associated disorder classification and clinical score regression tasks,
using cross-entropy loss and mean absolute error loss to fine-tune this task-specific model, respectively.
To enhance interpretability of TGA, we incorporate a \emph{learnable attention mask} $M$ to automatically detect discriminative brain ROIs and functional connectivities. 
This is achieved by replacing the original {\small{$\hat{A} $}}  in Eq.~\eqref{Eq1} with 
{\small{$\hat{A} \mathop{\vcenter{\hbox{$\scriptstyle \bigotimes$}}} M$}}, where  {\small{$\mathop{\vcenter{\hbox{$\scriptstyle \bigotimes$}}}$}} is element-wise multiplication.

\begin{table}[!t]
\setlength{\abovecaptionskip}{-1pt} 
\setlength{\belowcaptionskip}{-1pt}  
\setlength\abovedisplayskip{-1pt}
\setlength\belowdisplayskip{-1pt}
\renewcommand\tabcolsep{2.0pt}
\renewcommand{\arraystretch}{0.8}
\scriptsize
\centering
\caption{Results (\%) of eight methods in the task of ANI vs. HC classification on the HAND dataset.}
\resizebox{\columnwidth}{!}{
\begin{tabular}{l|c|c|c|c|c}
\toprule

Method & AUC~$\uparrow$ & ACC~$\uparrow$ & SEN~$\uparrow$ & SPE~$\uparrow$ & BAC~$\uparrow$ \\
\midrule

        SVM&61.7(3.2)&57.3(2.7)	&58.3(4.3)&57.2(2.9)&57.7(2.7) \\ 
        XGBoost&56.7(3.6)&53.3(2.8)&51.7(3.6)&56.7(6.6)&54.2(2.8)\\
        GCN &64.2(2.1)&59.6(2.3)&58.7(5.1)&61.7(4.9)&60.2(2.4)\\
       GAT &65.7(3.9)&61.1(3.0)&\textbf{62.5(5.8)}&59.5(6.0)&61.0(0.1)\\
      BrainNetCNN &65.4(1.6)&61.9(1.4)&61.7(3.7)&63.7(2.9)&62.7(1.1)\\
       BrainGNN &63.0(3.2)&60.3(2.6)&56.3(6.0)&63.7(5.6)&	60.0(1.8)\\
       TGAN~(Ours) &67.1(2.1)&60.5(3.8)&60.9(6.6)&60.9(6.4)&60.9(4.2)

       \\
TGAE~(Ours) &\textbf{69.9(3.5)}&\textbf{62.4(3.4)}&60.2(5.7)&\textbf{65.3(9.1)}&\textbf{62.8(3.2)}

 \\
		\bottomrule
	\end{tabular}
 }
\label{main_classification}
\end{table}


\section{Experiments}
\label{sec:illust}

\vspace{-2pt}
\subsection{Experimental Setting} 
When using the proposed HND and WER strategies for graph augmentation, our methods are called \textbf{TAGN} and  \textbf{TAGE}, respectively.
In the experiments, we compare our methods with
two traditional methods (\ie, \textbf{SVM/SVR} 
and \textbf{XGBoost}
) that use 464-dimensional node statistics as input,
and four state-of-the-art (SOTA) deep learning methods (\ie, \textbf{GCN}~\cite{kipf2016semi},
\textbf{GAT}~\cite{velivckovic2017graph},
\textbf{BrainNetCNN} ~\cite{kawahara2017brainnetcnn}, 
and \textbf{BrainGNN}~\cite{li2021braingnn}
). 

\if false
, including 

2 typical graph neural networks (\ie, \textbf{GCN}~\cite{kipf2016semi} and \textbf{GAT}~\cite{velivckovic2017graph}) 
and 2 methods specifically designed for brain network analysis (\ie, \textbf{BrainNetCNN}~\cite{kawahara2017brainnetcnn} 
 and \textbf{BrainGNN}~\cite{li2021braingnn}). 
\fi

A 5-fold cross-validation strategy is 
used. 
In each fold, $80\%$ of data is used for fine-tuning 
while $20\%$ of data is used for validation. 
Five metrics are used for classification: area under ROC curve (AUC), accuracy (ACC), sensitivity (SEN), specificity (SPE), and balanced accuracy (BAC). 
Three metrics are used for regression: mean absolute error (MAE), mean squared error (MSE), and PC coefficient (PCC).

\if false
Our method is implemented in Pytorch. 
The Adam optimizer and two GCN layers with hidden dimension of $64$ are used in the two models.
The node dropping ratio $\alpha$ and edge removing ratio $\beta$
are empirically set as $10\%$ and $50\%$, respectively. %
\fi

\vspace{-2pt}
\subsection{Result and Analysis} 
\vspace{-4pt}
\noindent \textbf{(1) Classification and Regression Results}. 
We report the results achieved by eight methods in  classification and regression tasks in Table~\ref{main_classification} and Table~\ref{main_regression}, respectively. 
It can be observed from Table~\ref{main_classification} that our methods generally show notable superiority over the traditional SVM
and XGBoost methods in the classification task.
The AUC of our TGAE is $13.2\%$ higher than XGBoost in distinguishing ANI from HC.
Compared with four SOTA methods (\eg, BrainGNN),
our methods yield better performance in most metrics. 
In addition, we can see from Table~\ref{main_regression}, our methods generally 
outperform six competing methods in cognitive score regression. 
These findings 
imply that our methods can learn discriminative graph representations in the two downstream tasks.

\begin{table}[!t]
\renewcommand{\arraystretch}{0.8}
\setlength{\abovecaptionskip}{-1pt} 
\setlength{\belowcaptionskip}{-1pt}  
\setlength\abovedisplayskip{-1pt}
\setlength\belowdisplayskip{-1pt}
\renewcommand\tabcolsep{2.0pt}
\scriptsize
\centering
\caption{Regression results achieved by eight methods for predicting attention/working memory scores on HAND.}
\resizebox{\columnwidth}{!}{
\begin{tabular}{l|c|c|c}
\toprule
Method  & MAE $\downarrow$& MSE $\downarrow$& PCC $\uparrow$  \\
 
 
		\midrule
SVR &0.1767$\pm$0.0081 &0.0485$\pm$0.0039& 0.1091$\pm$0.0297\\
XGBoost &0.1511$\pm$0.0034&0.0381$\pm$0.0019&  0.1340$\pm$0.0245\\
GCN&0.1577$\pm$0.0069&0.0421$\pm$0.0033& 0.1502$\pm$0.0481

                 \\
       GAT &0.1556$\pm$0.0010&0.0410$\pm$0.0006&0.0920$\pm$0.0238\\
       BrainNetCNN & 0.1647$\pm$0.0091& 0.0457$\pm$0.0050&0.1451$\pm$0.0298\\

       BrainGNN&0.1610$\pm$0.0022&0.0418$\pm$0.0010& 0.1413$\pm$0.0500 \\

             TGAN~(Ours) &0.1477$\pm$0.0031&0.0373$\pm$0.0011&0.1819$\pm$0.0527
             \\
TGAE~(Ours)  &\textbf{0.1464$\pm$0.0020}&\textbf{0.0363$\pm$0.0011}&\textbf{0.1952$\pm$0.0297}

\\

		\bottomrule
	\end{tabular}
 }
\label{main_regression}
\end{table}

\begin{figure*}[!t] 
\setlength{\abovecaptionskip}{0pt}
\setlength{\belowcaptionskip}{0pt}
\setlength{\abovedisplayskip}{0pt}
\setlength\belowdisplayskip{0pt}
	\centering
	\includegraphics[width=0.98\linewidth]{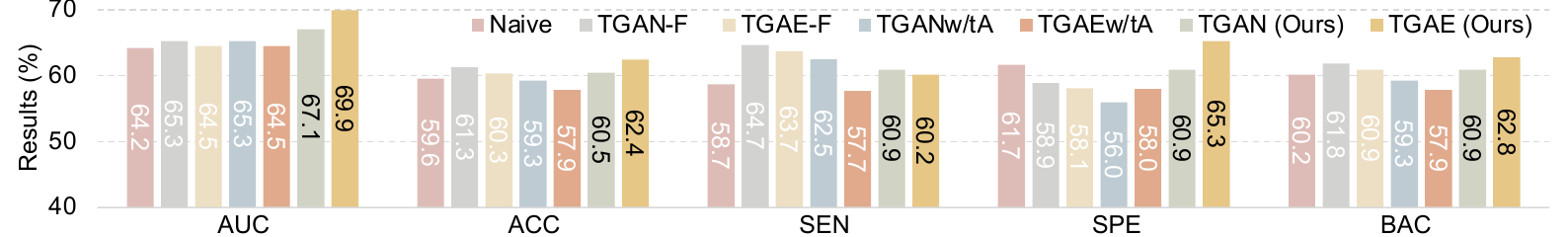}
	\caption{Results of our methods 
 and its variants in ANI vs. HC classification.}
	\label{ablation}
\end{figure*}

\vspace{2pt}
\noindent \textbf{(2) Ablation Study}.  
We compare TGAN and TGAE with 5 variants:
 \textbf{Naive} trained on the target cohort without the pretext model,
\textbf{TGAN-F} and \textbf{TGAE-F} that directly apply the pretrained GCN encoder to the task-specific model without fine-tuning, 
\textbf{TGANw/tA} and \textbf{TGAEw/tA} 
without the learnable attention mask, with results 
reported in Fig.~\ref{ablation}.
From Fig.~\ref{ablation}, TGAN and TGAE are superior to the Naive, 
validating the necessity of utilizing auxiliary fMRI 
for
model pretraining 
when dealing with
small-scale target data. 
Besides, our methods 
generally outperform frozen counterparts (\ie, TGAN-F and TGAE-F), 
indicating that fine-tuning can enhance the adaptability of the pretrained encoder to target data. 
Additionally, our methods generally outperform TGANw/tA and TGAEw/tA in most cases, 
suggesting the effectiveness of the attention mask in learning discriminative graph features.

\begin{figure}[!t] 
\setlength{\abovecaptionskip}{0pt}
\setlength{\belowcaptionskip}{0pt}
\setlength{\abovedisplayskip}{0pt}
\setlength{\belowdisplayskip}{0pt}
	\centering
	\includegraphics[width=1\linewidth]{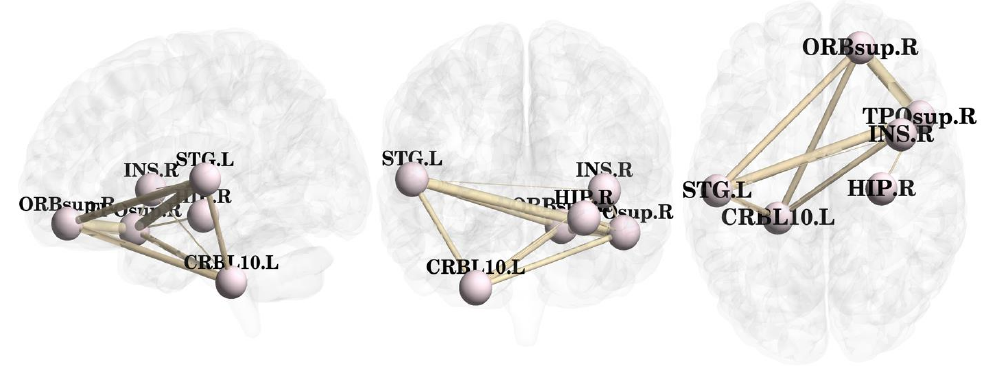}
	\caption{Top 10 most discriminative functional connectivities   in the ANI vs. HC
classification task.}
	\label{biomarker}
\end{figure}

\vspace{2pt}
\noindent \textbf{(3) Discriminative Functional Connectivity}.  
In Fig.~\ref{biomarker}, we visualize the top 10 discriminative functional connectivities (FCs) identified by our method (with line width denoting attention weight) in ANI vs. HC classification. 
Figure~\ref{biomarker} suggests that 
several ROIs are commonly detected in the identified FCs, including  
\emph{orbital part of superior frontal gyrus}, 
\emph{temporal pole of superior temporal gyrus},  and \emph{superior temporal gyrus}.  
This aligns with previous HIV-associated observations. 
These results demonstrate that the FCs and ROIs detected by our method have good interpretability, which can be used as 
potential imaging biomarkers for early identification
of HIV-associated neurological changes.

\section{CONCLUSION}
This work presents a topology-aware graph augmentation (TGA) framework for fMRI analysis. 
The TGA comprises a pretext model trained on auxiliary unlabeled fMRI  cohorts and a task-specific model fine-tuned on a target dataset. 
Two novel topology-aware graph augmentations are designed 
to facilitate self-supervised training. 
Experiments suggest the effectiveness of TGA in classification and regression tasks. 


\if false
\section{COMPLIANCE WITH ETHICAL STANDARDS}
This is a numerical simulation study for which no ethical approval was required.

\section{ACKNOWLEDGMENTS}
The research of Q.~Wang, A.~Bozoki and M.~Liu was supported in part by NIH grants (Nos.~AG073297 and AG082938). 
\fi


\bibliographystyle{IEEEbib}
\bibliography{mybib}

\end{document}

%% file: defs.tex
\def\0{{\bf 0}}
\def\1{{\bf 1}}

\def\eg{{\em e.g.}}
\def\ie{{\em i.e.}}